\documentclass[ApJ,]{aastex701}
\usepackage{amsmath,amsfonts,amssymb,cancel}
\usepackage{fix-cm}
\graphicspath{{./figures/}}

\renewcommand{\vec}[1]{ \protect {{\mathbf{\boldsymbol{#1}}}}}

\begin{document}


\title{Ion-scale Turbulence and Energy Cascade Rate in the Solar Corona and Inner Heliosphere}

\author[0000-0002-8078-0902]{Eduard P. Kontar}
\affiliation{School of Physics \& Astronomy, University of Glasgow, Glasgow, G12 8QQ, UK}
\email[]{eduard@astro.gla.ac.uk} 

\author[0000-0001-8720-0723]{A. Gordon Emslie}
\affiliation{Department of Physics \& Astronomy, Western Kentucky University, Bowling Green, KY 42101, USA}
\email[show]{gordon.emslie@wku.edu} 

\author[0000-0003-1967-5078]{Daniel L. Clarkson}
\affiliation{School of Physics \& Astronomy, University of Glasgow, Glasgow, G12 8QQ, UK}
\email[show]{daniel.clarkson@glasgow.ac.uk} 

\author[0000-0001-8913-191X]{Alexander Pitňa}
\affiliation{Charles University, Faculty of Mathematics and Physics, Prague 8, Czech Republic}
\email[show]{alexander.pitna@mff.cuni.cz} 

\begin{abstract}

Plasma turbulence cascading from MHD to kinetic scales in the heliospheric plasma is believed to play a key role in coronal heating and fast solar wind acceleration, but the properties of the turbulence remain poorly constrained by observations. Here we compare the ion-scale density fluctuation levels inferred from the properties of solar radio bursts with the magnetic field fluctuation levels obtained through in-situ measurements in the inner heliosphere. We find that the observed magnetic and density fluctuation amplitudes are consistent with excitation by kinetic Alfv\'en waves and/or KAW structures over broad range of distances from the Sun. We then use the radio diagnostics and the KAW scenario to deduce the radial variation of magnetic fluctuation amplitudes in regions close to the Sun where in-situ measurements cannot be obtained. Further, we calculate the energy cascade rate (plasma heating rate) profile over a region that extends from the low corona ($\sim 0.1$~R$_\odot$) into the heliosphere (out to $\sim 1$~au), and compare it to the energy deposition rate required to drive the solar wind. The cascade rate agrees with the available in-situ measurements and also provides predictions closer than $\sim 10$~R$_\odot$ where in-situ approaches are not available. The results provide unique diagnostics of the ion-scale plasma turbulence amplitude and energy cascade rate spanning over three orders of magnitude in solar distance.

\end{abstract}


\section{Introduction}

The solar corona and heliosphere is a turbulent plasma medium, in which waves and/or nonlinear structures are believed to play a key role in transferring energy and momentum to particles. 
In particular, cascading MHD turbulence is believed to be an essential element in heating the solar corona and accelerating the solar wind \citep[see, e.g.,][for recent reviews]{2013LRSP...10....2B,2015RSPTA.37340148C}. However, the properties of the turbulence, and its radial evolution at the ion-scales where dissipation of energy is anticipated, 
are poorly understood, largely due to a general lack of observational constraints, 
both on the large MHD scales and on kinetic scales 
\citep[e.g.,][]{2013LRSP...10....2B,2015RSPTA.37340148C,2015RSPTA.37340147G}. 
Accordingly, the precise mechanisms that heat the corona, and accelerate and heat the solar wind, remain largely open questions \citep[e.g.,][]{2015RSPTA.37340269D,2015RSPTA.37340155K,2024FrASS..1171058S}.

The properties of MHD turbulence in the heliosphere can be inferred through both in-situ measurements and remote observations. Both approaches enrich our overall understanding of the underlying turbulence in their own way, providing insight into the nature of the solar coronal turbulence. In-situ observations suggest that large-scale Alfv\'en waves \citep{1971JGR....76.3534B} cascade \citep{1995SSRv...73....1T}, with little or no dissipation, to smaller (ion) scales, where the energy in the turbulence is partially transferred to ambient ions and partially cascades further towards even smaller (electron) scales.

Observations of density fluctuations in the corona and the solar wind 
\citep[e.g.,][]{1990ApJ...358..685A,2005JGRA..110.3101H,2009ApJ...707.1668C,2017ApJ...850..129S,2021NatAs...5..796R,2023ApJ...956..112K}, 
suggests that the turbulence in solar plasma is compressive. This introduces challenges 
and complications for a theoretical description 
\citep[e.g.,][]{1993PhFlA...5..257Z,2015RSPTA.37340147G} but also provides an opportunity 
for density-fluctuation based diagnostics to be used in assessing the properties of the turbulence. Specifically, variations of the plasma refractive index associated with density inhomogeneities lead 
to both elastic and inelastic scattering of radio waves, 
particularly for waves produced by plasma processes at frequencies 
close to the plasma frequency. As a result, radio observations, both of solar bursts 
and of celestial sources that are closely aligned with the Sun,
provide a unique diagnostic of the compressive fluctuations 
and waves in the solar corona and heliosphere at ion scales 
\citep[e.g.,][]{2005JGRA..110.3101H,2009ApJ...707.1668C}. 
Recent analysis of (mostly) solar radio bursts 
\citep[][]{2019ApJ...884..122K,2023ApJ...956..112K,2024ApJ...968...72A} has provided information about anisotropic density fluctuations over a range of solar distances 
from the low corona to 1~au, complementing recent in-situ observations in the inner heliosphere from Parker Solar Probe at distances $>0.1$~au \citep[PSP;][]{2016SSRv..204....7F}.

High-time-resolution solar wind observations, mostly near 1~au, suggest that at ion-scales both the magnetic and density fluctuations are consistent with quasi-perpendicular kinetic Alfv\'en waves
\citep[e.g.,][]{2013PhRvL.110v5002C,2017JGRA..122.6940R,2019ApJ...879...82P,2019PhRvX...9c1037G},
which can play a key role in the energy cascade and subsequent dissipation of turbulent energy in the heliosphere. In situ magnetic field and (in-situ plus remote sensing) density fluctuation data can therefore be used in combination to test the validity of the kinetic Alfv\'en wave excitation hypothesis, to constrain its amplitude, and to determine the associated energy cascade rate, even at locations that are inaccessible to in-situ spacecraft measurements.

In Section~\ref{sec:in-situ-observations} we briefly review and exemplify in-situ observations of magnetic field, velocity and density fluctuation spectra near ion-scales. 
In Section~\ref{sec:radio-observations} we discuss the inference of magnetic fluctuation amplitudes 
through radio observations, and we show that, overall, the inferred behaviors of magnetic field 
fluctuations with heliocentric distance are consistent with excitation by kinetic Alfv\'en waves. 
In Section~\ref{sec:heating-rate} we use radio data to estimate the energy cascade rate in the corona for fast and slow solar wind conditions, 
and hence to determine the heating rate associated with damping of the turbulent fluctuations, which are then compared with in-situ observations 
at distances $>10$~R$_\odot$. In Section~\ref{sec:summary} 
we summarize the results obtained.

\section{In-situ observations of Kinetic Alfv\'en waves}\label{sec:in-situ-observations}

\begin{figure}[pht!]
\centering
\includegraphics[width=0.45\textwidth]{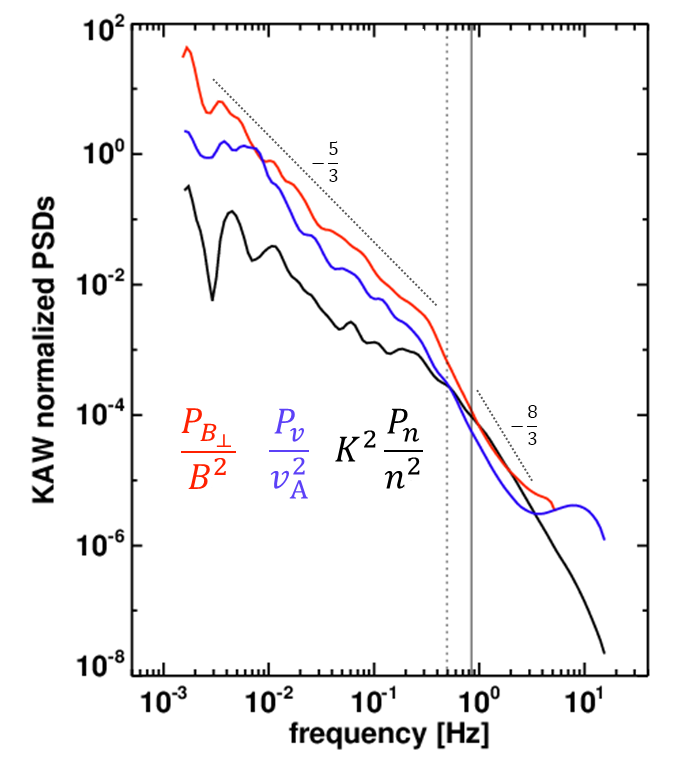}
\includegraphics[width=0.49\textwidth]{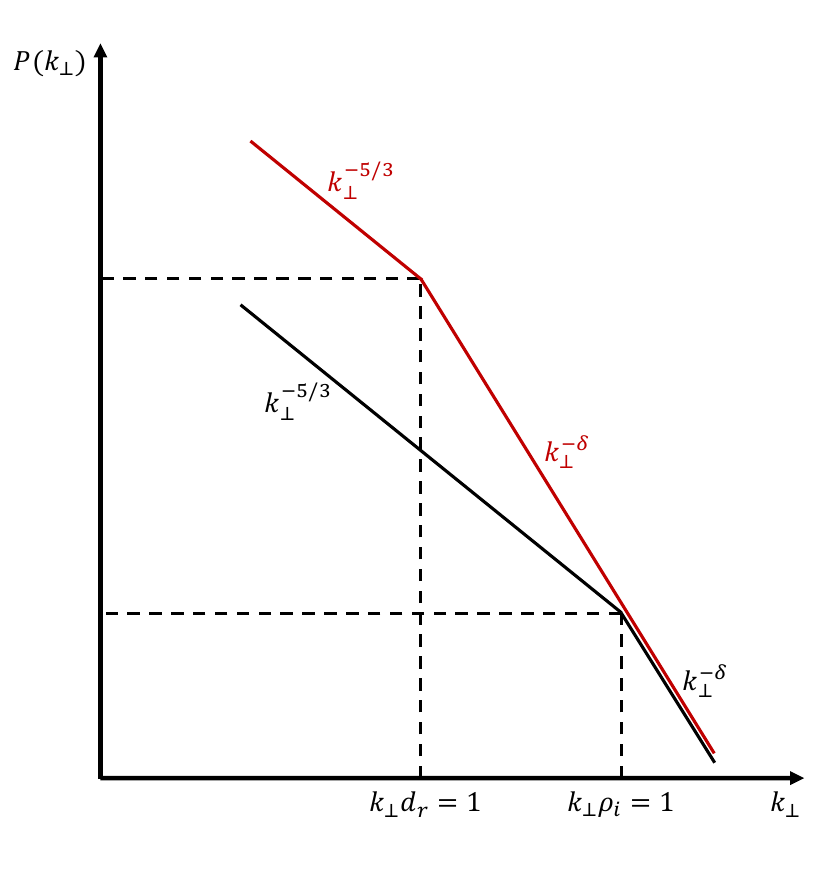}
\caption{\textbf{Left:} KAW normalized PSDs of magnetic field (red), velocity (blue), and ion density (black) fluctuations (computed using $K^2=\beta_i(1+\beta_i)$), for an event observed by both Spektr-R and Wind on 2012~April~27 at approximately 21:00~UT. The gray dotted lines show the analytic approximations to the spectrum, with a $k^{-5/3}$ power-law form in the inertial range, and a steeper $k^{-8/3}$ form in the dissipative range. The dashed and solid vertical gray lines show the approximate spacecraft-frame frequencies $f_{d_i}$ and $f_{\rho_i}$ respectively (see Table~\ref{table1}). \textbf{Right:} Model wavenumber spectra of the magnetic field (red) and ion density (black) fluctuations. The spectrum of magnetic fluctuations breaks from the $k_\perp^{-5/3}$ Kolmogorov form of the inertial range to the steeper $k_\perp^{-\delta}$ form (with $\delta \simeq 8/3$) in the dissipative range, at a wavenumber $d_r^{-1}$. The density fluctuations exhibit a similar break, but at a higher wavenumber corresponding to the thermal proton gyroradius $\rho_i$.
}
\label{fig:obs_cartoon}
\end{figure}

Both observations and simulations of solar wind turbulence at ion scales support the presence of kinetic Alfv\'en waves (KAW) and/or coherent KAW-like structures on kinetic scales 
\citep{2013PhRvL.110v5002C,2017JGRA..122.6940R,2018ApJ...853...26F,2019ApJ...870..106W,2019ApJ...879...82P,2019PhRvX...9c1037G,2021ApJ...917L..12P}. The in-situ observed spectrum of magnetic fluctuations 
is normally close to $k^{-5/3}$ over the inertial range encompassing MHD scales, 
with a steeper spectrum $k^{-\delta}$ observed at ion-scales, 
with a spread of $\delta$ 
values from about $2 - 4$, clustering around a mean value $\delta \simeq 8/3$ 
\citep[][see the example spectrum in Figure~\ref{fig:obs_cartoon}]{1998JGR...103.4775L,2006ApJ...645L..85S,2009PhRvL.103p5003A,2012ApJ...760..121A,2017ApJ...850..120R,2022JGRA..12729483R}. The density fluctuation spectrum is also observed to be close to $k^{-5/3}$ 
in the inertial range \citep[e.g.,][]{1989ApJ...337.1023C,1990JGR....9511945M,2018ApJ...856...73C}.
For density fluctuations the break between the inertial and kinetic ranges occurs close to the ion gyroradius $\rho_i=v_{Ti}/\omega_{ci}$ \citep[e.g.,][]{2016ApJ...825..121S,2019ApJ...872...77S}, whereas for the magnetic field fluctuations, the break is at $k \simeq d_r^{-1}=(d_i+\rho_i)^{-1}$ \citep[e.g.,][]{2014ApJ...787L..24B,2023ApJ...942...93L}, 
effectively the smaller value of the proton inertial length $d_i=c/\omega_{pi}$ 
and the proton gyro-radius \citep{2014GeoRL..41.8081C}. Here, $c$ (cm~s$^{-1}$) is the speed of light, $\omega_{pi} = \sqrt{4 \pi \, n \, e^2/m_i}$ (s$^{-1}$) is the ion plasma frequency, with $e$ (esu) the proton charge, $m_i$ (g) the ion mass, and $n$ (cm$^{-3}$) is the ambient density. Also, $v_{Ti}=\sqrt{2k_BT_i/m_i}$ (cm~s$^{-1}$) is the ion thermal speed, with $k_B$ (erg~K$^{-1}$) being Boltzmann's constant and $T_i$ (K) the ion temperature, and $\omega_{ci} = e B/m_i c$ is the ion gyrofrequency, with $B$ (G) being the magnetic field strength. The ratio of the two pertinent scales is $\rho_i/d_i = \sqrt{8 \pi \, n \, k_B \, T_i/B^2} \equiv \sqrt{\beta_i}$, where the ion plasma beta $\beta_i$ is the ratio of the energy density in ions to the energy density in the magnetic field.

\begin{figure}
 \centering
\includegraphics[width=0.6\textwidth]{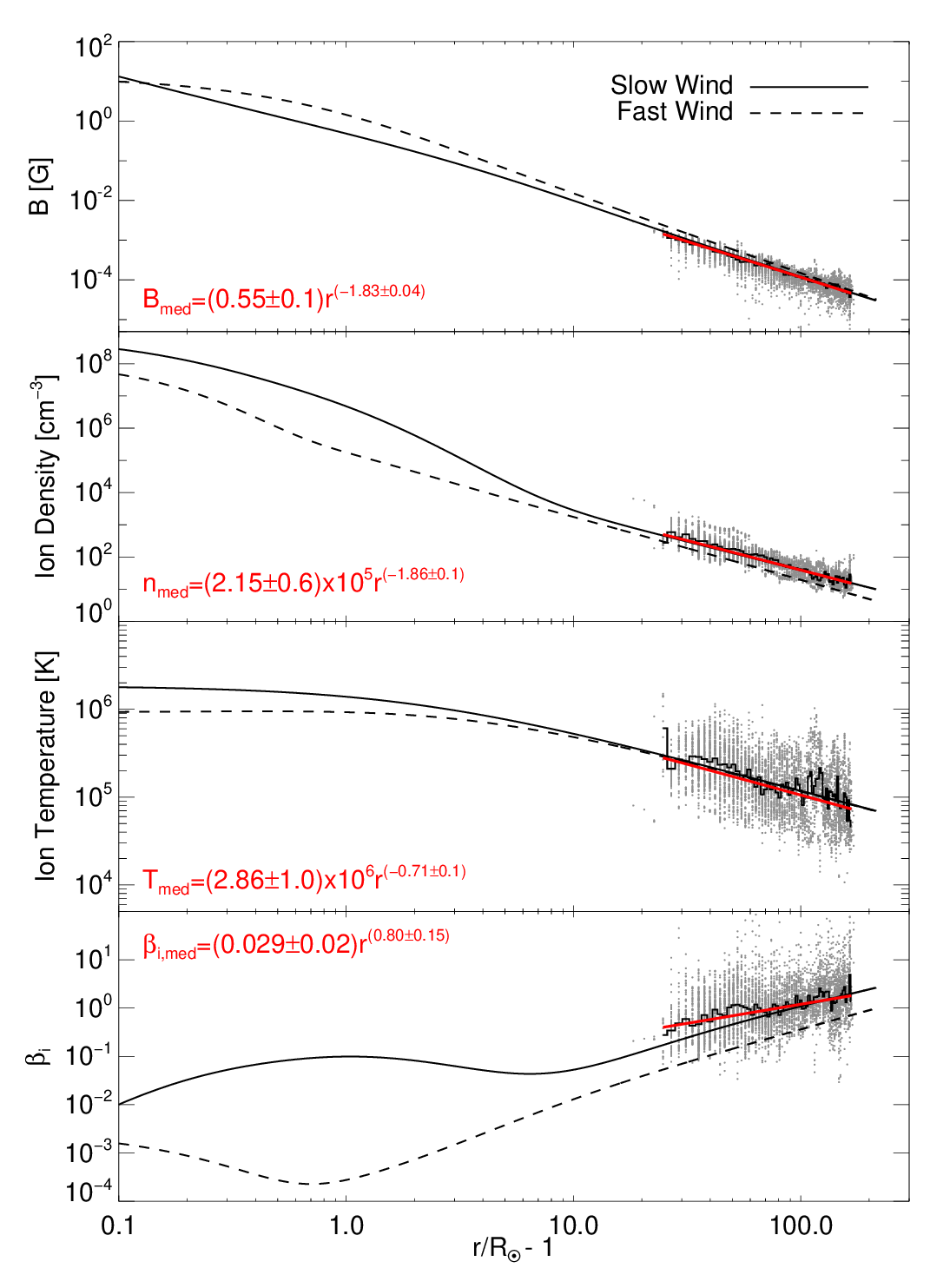}
\caption{\label{fig:psp_plasma} Radial variation of various plasma parameters, as observed by the Parker Solar Probe covering the time between between 2018-Oct-31 and 2023-Jul-31. From top to bottom: magnetic field strength $B$ (G), ambient ion density $n$ (cm$^{-3}$), ion temperature $T_i$ (K), and the ion plasma beta: $\beta_i = 8 \pi \, n \, k_B \, T_i/B^2$. In each plot power-law fits (red) to the median observed values (black histograms) are shown. The solid and dashed lines show the analytic models provided in the Appendix, for the slow and fast solar wind, respectively.}
\end{figure}

Figure~\ref{fig:psp_plasma} shows the variation of several plasma parameters of interest, as a function of distance from the surface of Sun (in units of the solar radius), together with the best-fit power-law expressions to the median values. From top to bottom are the magnetic field strength (G), the ion density (cm$^{-3}$), the ion temperature (K), and the ion plasma beta $\beta_i = \sqrt{8 \pi \, n \, k_B \, T_i/B^2}$. From the bottom panel of Figure~\ref{fig:psp_plasma}, we see that in most regions of the inner heliosphere, $\beta_i < 1$ , so that $d_i > \rho_i$. For such conditions the ion-scale break $k_r=d_r^{-1}=(d_i+\rho_i)^{-1}$ for magnetic field fluctuations occurs at $k \simeq d_i^{-1}$, a lower wavenumber than the break at $k \simeq \rho_i^{-1}$ in the density fluctuation wavenumber spectrum.

\begin{table}[pht]
\caption{Values of parameters within the interval analyzed in Figure~\ref{fig:obs_cartoon}}\label{table1}
\centering
\begin{tabular}{ c c }
\hline
Start time (Wind) & 2012-04-27 20:38 \\ 
 End time (Wind) & 2012-04-27 21:08 \\
 Start time (Spektr-R) & 2012-04-27 21:32 \\ 
 End time (Spektr-R) & 2012-04-27 22:02 \\
 Bulk solar wind speed $V_{sw}$ & {470 km~s$^{-1}$} \\
 Magnetic field-solar wind angle, $\theta_{VB}$ & $80^{\circ}$ \\ 
 Normalized residual energy $\sigma_{\rm r}$ & -0.62 \\
 Normalized cross-helicity $\sigma_{\rm c}$ & -0.10 \\
$f_{d_{\rm i}} = V_{sw}/2\pi d_i$ & $0.48$ Hz \\
$f_{\rho_{\rm i}} = V_{sw}/2 \pi \rho_i$ & $0.84$ Hz \\
$\beta_{\rm i}$ & $0.34$ \\
$K = \sqrt{\beta_i \, (1 + \beta_i)}$ & $0.67$ \\
\hline
\end{tabular}
\end{table}

The right panel of Figure~\ref{fig:obs_cartoon} shows the characteristic wavenumber spectra schematically, with the magnetic field fluctuation power spectrum in red and the density fluctuation power spectrum in black. The left panel of that Figure shows an example of how these dependencies are evident in the spectral analysis of single spacecraft measurements of density and magnetic field, under a Taylor hypothesis that assumes the turbulent eddies to be frozen into the mean solar wind flow, so that the measured fluctuation frequency $f$ at a fixed point in space is related to the turbulent wavenumber $k$ by $f = k V_{sw}/2\pi$, where $V_{sw}$ is the local bulk solar wind speed.  Table~\ref{table1} provides some details about the event in question, which occurred on 2012~April~27 at approximately 21:00~UT, including the time intervals of the event as observed by both spacecraft, the angle between the solar wind velocity and the magnetic field, the normalized residual energy $\sigma_r$ and cross-helicity $\sigma_c$ \citep[calculated as $\sigma_{\rm r} = \left ( \left < \mathbf{v}^2 \right > - \left < \mathbf{b}^2 \right > \right ) / \left(\left < \mathbf{v}^2 \right > + \left < \mathbf{b}^2 \right >\right)$ and $\sigma_{\rm c} = \left ( \left < \mathbf{z}_+^2 \right > - \left < \mathbf{z}_-^2 \right > \right ) / \left(\left < \mathbf{z}_+^2 \right > + \left < \mathbf{z}_-^2 \right >\right) = 2 \left < \mathbf{v} \cdot \mathbf{b} \right >/\left(\left < \mathbf{v}^2 \right > + \left < \mathbf{b}^2 \right >\right)$, where $(\mathbf{z}_+, \mathbf{z}_-) \equiv \mathbf{v} \pm \mathbf{b}$ are the Elsasser variables; see Equations~(1) and (2) of][]{2013ApJ...778..177W}, 
the key break frequencies $f_{d_{\rm i}}$ and $f_{\rho_{\rm i}}$ in the observed spectra, 
and the values of the ion plasma beta $\beta_i$ and related quantity $K$ (Equation~\eqref{eq:dn2_dB2_kin_alfven}).

Shown are the wavelet power spectra of the magnetic field obtained by the MFI magnetometer instrument \citep{1995SSRv...71..207L} onboard the Wind spacecraft situated near the inner L1 Lagrangian point, 
and the density spectrum computed from high-cadence observations of ion density by the Bright Monitor of the Solar Wind (BMSW) instrument \citep{2013SSRv..175..165S} onboard the Spektr-R \citep{2012SoSyR..46..466A} spacecraft situated in Earth orbit. 
First, power spectral densities (PSDs) are estimated via a continuous wavelet transform (CWT) \citep{1998BAMS...79...61T}, using a Morlet mother wavelet with center frequency $\omega = 6$~s$^{-1}$, obtaining the trace PSD of magnetic field fluctuations $P_B = P_{B_{\rm x}} + P_{B_{\rm y}} + P_{B_{\rm z}}$, where $P_{B_{\rm x}}$, $P_{B_{\rm y}}$ and $P_{B_{\rm z}}$ are individual PSDs of the magnetic field components. Next, a PSD of the magnitude of the magnetic field, $P_{|B|}$, is estimated. Finally, the PSD of perpendicular magnetic field fluctuations is computed as $P_{B_{\perp}} = P_B - P_{|B|}$. For density fluctuations, we denote the CWT PSD as $P_n$. The figure shows the KAW normalized PSDs, calculated as $P^{\rm KAW}_{P_{B_{\perp}}} = P_{B_{\perp}}/B_0^2$, $P^{\rm KAW}=P_v/v_A^2$, and $P^{\rm KAW}_n = K^2 P_{n}/n^2$, where $K^2 = \beta_i (1 + \beta_i)$ (see Equation~\eqref{eq:dn2_dB2_kin_alfven}). The velocity fluctuation spectrum (blue) closely matches the magnetic field fluctuation spectrum (red), as expected for an Alv\'enic perturbation. One can see from Figure~\ref{fig:obs_cartoon} that above the spacecraft frame frequency $f_{\rho_{i}}= V_{sw}/2\pi\rho_{i}$ corresponding to the proton thermal gyroradius, the normalized spectra of the density and magnetic field fluctuations (left panel) indeed exhibit forms that are very similar to the idealized analytical wavenumber spectra shown in the right panel.

The polarization properties of linear KAWs establish characteristic relations that link fluctuations in the plasma density $\delta n$, the electric 
\citep[e.g.,][]{2010PhPl...17f2308B} and magnetic field $\delta B_\perp$, 
and the bulk flow velocity $\delta v$. In the limit $k_{\perp}\gg k_{\parallel}$, representing a wave propagation direction that is nearly perpendicular to the magnetic field, and for perpendicular wave numbers in the range $\omega_{ci}/v_{Ti} \equiv 1/\rho_{i} < k_{\perp} < 1/\rho_{\rm e} \equiv \omega_{ce}/v_{Te}$, the relation between the normalized amplitude of density $\delta n/n$ and normalized perpendicular magnetic field $\delta B_{\perp}/B$ fluctuations is \citep[e.g.,][]{2017JGRA..122.6940R,2018ApJ...853...26F,2019ApJ...879...82P,2019PhRvX...9c1037G}:

\begin{equation}\label{eq:dn2_dB2_kin_alfven}
 \frac{\delta B_\perp^2}{B^2} =\left(1+\frac{T_i}{T_e}\right) 
 \left(\frac{v_s}{v_A}\right)^2 \, \left[1+\left(1+\frac{T_i}{T_e}\right)\left(\frac{v_s}{v_A}\right)^2\right]\frac{\delta n^2}{n^2} = {\beta_i} \, (1+ \beta_i) \, \frac{\delta n^2}{n^2} \equiv K^2 \, \frac{\delta n^2}{n^2} \,\,\, .
\end{equation}
Here
$v_{s}=\sqrt{k_B T_e/m_i} $ (cm~s$^{-1}$) is the isothermal sound speed, 
with $T_e$ (K) the electron temperature, $v_A=B/\sqrt{4\pi \, m_i \, n}$ (cm~s$^{-1}$) is the Alfv\'en speed, and in the third equality an isothermal hydrogen plasma with $T_i \simeq T_e = T$ is assumed. The black line in the left panel of Figure~\ref{fig:obs_cartoon} shows the density spectrum $K^2 \, \delta n^2/n^2$ 
obtained from the measured values of $\beta_i$ (see Table~\ref{table1}). 
One can see that above $\rho_{i}^{-1}$, the normalized magnetic fluctuation spectrum 
(red line) and normalized density fluctuation spectrum (black line) 
are close to each other for $k\rho_i>1$, consistent with the KAW relation of Equation~\eqref{eq:dn2_dB2_kin_alfven}.

\section{Radial variations of density and magnetic fluctuations}\label{sec:radio-observations}

In this Section we review the inference of the density fluctuation profile from observations of radio sources near the plasma frequency. The right panel of Figure~\ref{fig:obs_cartoon}, which closely approximates the observed spectrum in the left panel, shows that the density wavenumber spectrum $S(\vec{k})$ is well approximated by a double power-law:
 
\begin{equation}\label{eq:spectrum}
\begin{aligned}
S(\vec{k})=\frac{\delta n^2_{\rho_i} }{n^2}\frac{\rho_i^3}{4\pi} \times 
\begin{cases}
q^{- \frac{5}{3} - 2} , & q \leq 1 \\ 
q^{-\delta-2} , & q >1
\end{cases}
\end{aligned}
 \end{equation}
where $n$ is the ambient density, $q$ is the magnitude of the dimensionless wavenumber measure $\mathbf{q} = \rho_i (\vec{k}_\perp , k_\parallel/\alpha)$ 
and $\alpha$ is an anisotropy parameter \citep[see Equation~(5) of][]{2023ApJ...956..112K}. The corresponding one-dimensional spectrum $k^2 S(k)$ is thus $\propto k^{-5/3}$ 
for $k\rho_i<1$ and $\propto k^{-\delta}$ for $k\rho_i>1$ . 

The mean scattering rate (proportional to the reciprocal of the mean free path) of a radio wave in plasma with density fluctuations is determined by the spectrum of density fluctuations $S(\vec{k})$, and is proportional to the spectrum-weighted mean wavenumber $\overline{q}$ \citep[cf. the quantity $\overline{q \, \epsilon^2}$ in Equation~(14) of][noting the factor of $\rho_i$ difference in the definition of $q$]{2023ApJ...956..112K}, found by direct integration of Equation~\eqref{eq:spectrum}:

\begin{equation}\label{eq:qbar}
   \overline{q} = \int q \, S(\vec{k}) \, d^3{\vec k} = \alpha \, \left(3+\frac{1}{\delta-2}\right) \frac{\delta n^2 _{\rho_i} }{n^2} \,\,\, ,
\end{equation}
and this can be compared with the empirical profile, as a function of distance $r$ from the Sun, deduced from radio observations. The result is

\begin{equation}\label{eq:qeps2}
   \alpha \, \left(3+\frac{1}{\delta-2}\right) \frac{\delta n^2 _{\rho_i} }{n^2} 
= \alpha \, \rho_i \frac{2 \times 10^3}{R_\odot} \left(1-\frac{R_\odot}{r}\right)^{2.7}\left(\frac{R_\odot}{r}\right)^{0.7} \,\,\, .
\end{equation}
Equation~\eqref{eq:qeps2} allows us to use radio observations to determine the magnitude of the density fluctuations at the break point $k_\perp \rho_i =1$, as a function of solar distance $r$:

\begin{equation}\label{eq:dn2_n2}
\frac{\delta n^2_{\rho_i}}{n^2(r)}
=\frac{\rho_i} {4.5 \, R_\odot} \, 
2 \times 10^3 \left(1-\frac{R_\odot}{r}\right)^{2.7}\left(\frac{R_\odot}{r}\right)^{0.7} \,\,\, ,
\end{equation}
where we have used $\delta \simeq 8/3$ as a typical value \citep[e.g.,][]{2012ApJ...760..121A} of the spectral index at high wavenumbers (cf. left panel of Figure~\ref{fig:obs_cartoon}). We note that the anisotropy parameter $\alpha$ appears on both sides of Equation~\eqref{eq:qeps2}, so that the result of Equation~\eqref{eq:dn2_n2} is valid for any value of $\alpha$.

The spacecraft-measured spectrum in frequency space $P(f)$
 is closely related to the wavenumber spectrum $S(\vec{k})$, 
 $P(f)=n^2 \int S(\vec{k}) \delta\left(f-\frac{\vec{k} \cdot \vec{V}_{\mathrm{SW}}}{2 \pi}\right) {d^3 k}$, 
 where $\vec{V}_{sw}$ is the solar wind velocity. Following Equation~(C5) of \cite{2023ApJ...956..112K}, we find the associated frequency spectrum at frequencies above the break frequencies ($f_{\rho_i}$ and $f_{d_r}$, respectively), the density and magnetic field fluctuation spectra are \citep[cf. Equation~(18) of][]{2009ApJ...696.1213P}

\begin{equation}\label{eq:density-spectrum-high-frequencies}
P_n(f) \, =
\frac{ \delta n^2 _{\rho_i}}{2 \, \delta f_{\rho_i}} \, 
\left ( \frac{f}{f_{\rho_i}} \right )^{-\delta} \, ; \qquad P_{B_\perp}(f) \, =
\frac{ B ^2 _{\perp d_r}} {2 \, \delta f_{d_r}} \, 
\left ( \frac{f}{f_{d_r}} \right )^{-\delta} \,\,\, .
\end{equation}
\begin{figure}
 \centering
\includegraphics[width=0.49\textwidth]{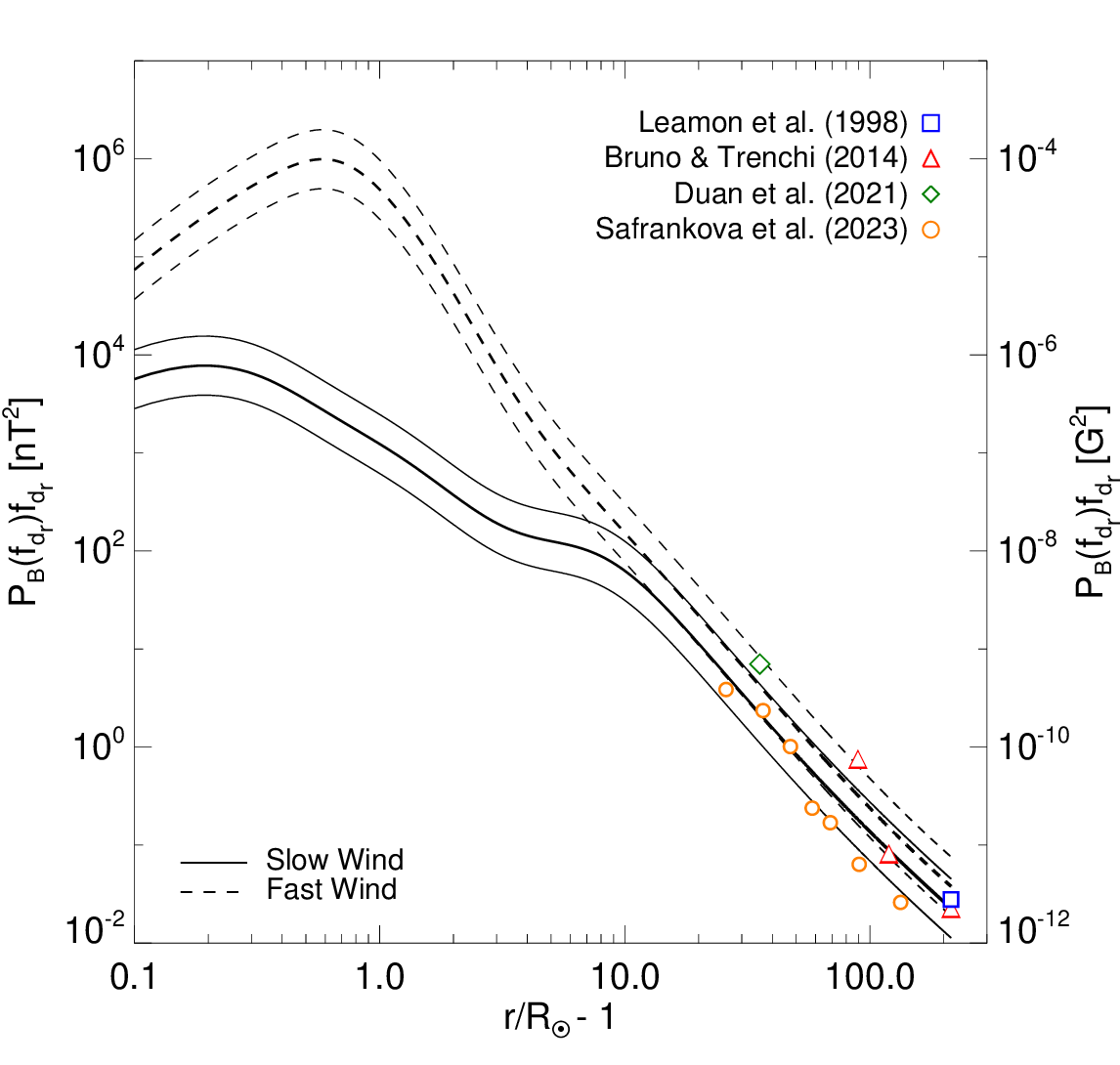}
\includegraphics[width=0.49\textwidth]{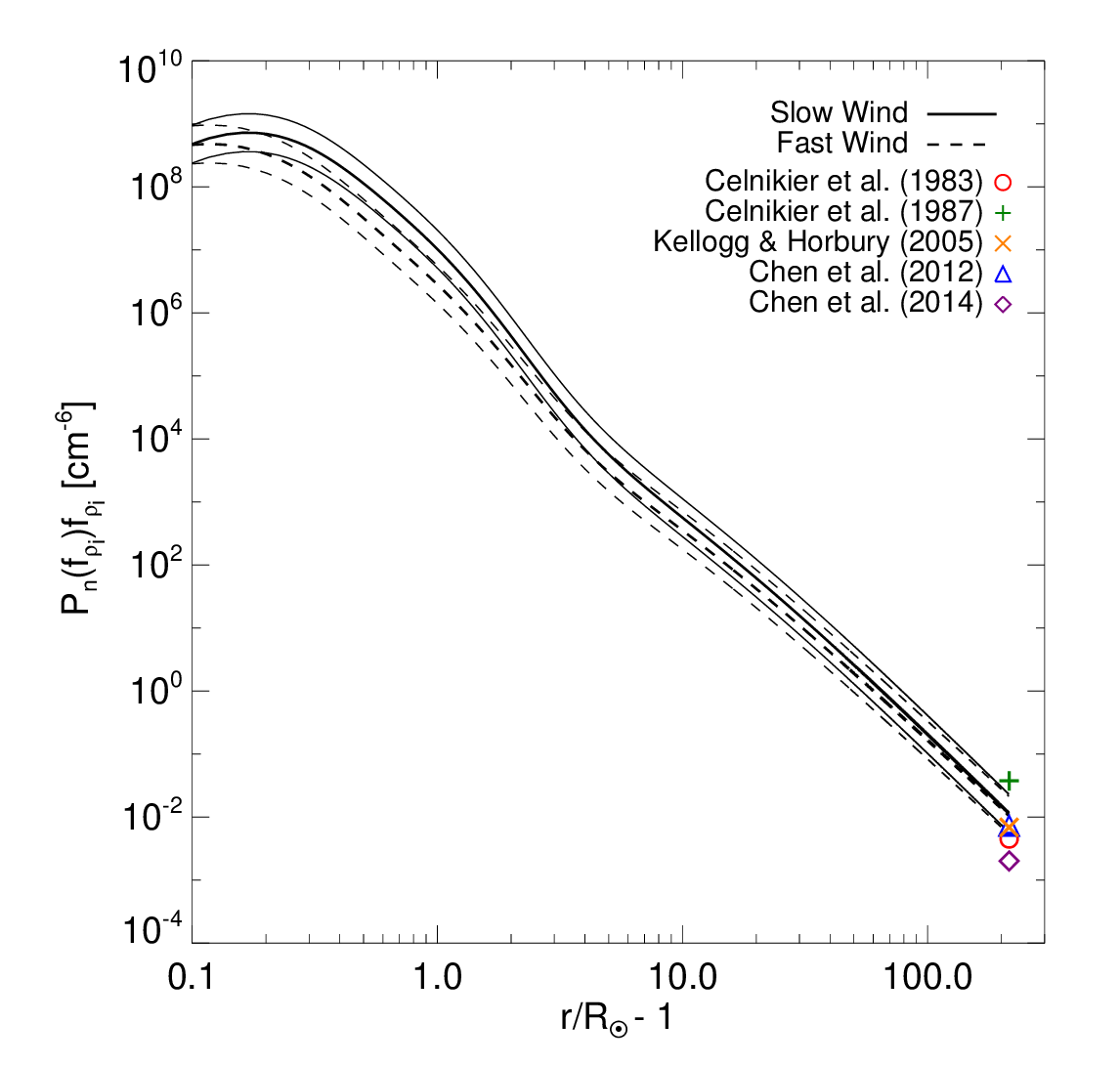}
\caption{\label{fig:p_B_break} \textbf{Left:} Magnetic fluctuations $P_B(f_{d_r}) \, f_{d_r}$ at the break and the predicted magnitude from radio-inferred density fluctuations. The solid and dashed lines show the expected values from kinetic Alfv\'en waves, for the slow and fast solar wind parameters given in the Appendix, and the plus and diamond symbols are based on measurements by PSP \citep{2021ApJ...915L...8D, 2023ApJ...946L..44S}. The bands represent a scaling factor from $0.5$ to $2$ that accounts for the spread in measurements \citep{2023ApJ...956..112K}. Additional data are from \cite{1998JGR...103.4775L, 2014ApJ...787L..24B}, as indicated. \textbf{Right:} Density fluctuations $P_n (f_{\rho_i}) \, f_{\rho_i}$ at the break, using Equation~\eqref{eq:density-spectrum-high-frequencies} with data at 1~au from \cite{1987A&A...181..138C, 2005AnGeo..23.3765K, 2014ApJ...789L...8C}, as shown in Figure~10 of \cite{2023ApJ...956..112K}.}
\end{figure}
The value of the magnetic field fluctuation energy density (i.e., the frequency-weighted power spectrum) at the break frequency $f_{d_r}$ is thus given by

\begin{equation}\label{eq:P_B_fB}
P_{B_\perp}(f_{d_r}) \, f_{d_r} =
\frac{ B^2 _{\perp d_r}}{2 \, \delta}
= \beta_i \, (1+\beta_i) \, \frac{B^2}{2 \, \delta} \, \left(\frac{d_r}{\rho_i}\right)^{\delta -1} \, 
\frac{\delta n^2 _{\rho_i}}{n^2} \, ,
\end{equation}
where in the second equality we have used Equation~\eqref{eq:dn2_dB2_kin_alfven} 
and the $k_\perp^{-\delta}$ form of the spectrum (right panel of Figure~\ref{fig:obs_cartoon}) 
to scale\footnote{The ratio 
$d_r/\rho_i=(v_{Ti}+v_A)/v_{Ti} = 1 + \sqrt{1/\beta_i}$ is close to unity at distances near the Earth (when $\beta_i \sim 1$) and is $\simeq \sqrt{1/\beta_i} \gg 1$ for $\beta_i \ll 1$, which is expected closer to the Sun and in the fast solar wind (Figure~\ref{fig:obs_cartoon}).} the magnetic fluctuation level at $k = \rho_i^{-1}$ to that at $d_r^{-1}$. 
In Figure \ref{fig:p_B_break} we show that the measured magnetic fluctuation levels 
compare well with the values obtained from the density fluctuation levels 
$ \delta n^2 _{\rho_i}/n^2$ using the radio-observation-inferred values 
of the magnetic fluctuations from Equation~\eqref{eq:P_B_fB}.

\section{Turbulence dissipation and the associated heating rate}\label{sec:heating-rate}

Measurements of the magnetic fluctuation amplitude $\delta B_\perp$ at the break (i.e., at the smallest scale in the inertial range) in the inner heliosphere also provide an opportunity to estimate the mean turbulent energy dissipation rate per unit mass (erg~g$^{-1}$~s$^{-1}$): $\varepsilon = - \, \partial E /\partial t$, where $E = \delta v^2/2 + \delta B_\perp^2/(8\pi \, m_i \, n)=Z^2/2$ is the energy (kinetic plus magnetic) of MHD turbulence per unit mass. Assuming Alfv\'enic fluctuations $\delta v/v_A\simeq \delta B_\perp/B$ and neglecting cross-helicity (cf. the low absolute value of $\sigma_c$ in Table~\ref{table1}), the energy cascade rate over the inertial range of the wavenumber spectrum $\varepsilon \propto Z_\lambda ^2 /\tau_\lambda$, where $Z_\lambda$ is the fluctuation amplitude at scale $\lambda$ 
and the nonlinear cascade time $\tau_\lambda=\lambda/Z_\lambda$. The resulting energy cascade rate is given by the approximate third-order law scaling law $\varepsilon = C \, Z_\lambda^3/\lambda$ \citep{1998GeoRL..25..273P,2023PhR..1006....1M}, where the constant $C={2 \, C_{\varepsilon}}/9 \sqrt{3}$ \citep[see ][for details]{2014ApJ...788...43U}, with a value $C_{\varepsilon}\simeq 0.22$ determined in the stationary case by \citet{2017PhRvE..95a3102L}. This gives $C\simeq 0.03$, similar to the value used in other works \citep[][]{2020ApJS..246...48B,2024ApJ...972L...8B} with which we compare our results.
When the energy proceeds only from large to smaller scales, i.e., there is no feedback from the small to larger scales, the energy transfer rate at the smallest scale of the inertial range equals the heating rate. We take the smallest scale in the inertial range to be $\lambda = d_r$, leading to an expression for the volumetric heating rate $Q(r)$ (erg~cm$^{-3}$~s$^{-1}$) for different heliocentric 
distances $r$:

\begin{equation}\label{eq:heating_rate}
 Q (r) =\varepsilon \, m_i \, n (r) = C \, m_i \, n(r) \ \frac{Z^3_{d_r} (r)}{d_r (r)} \,\,\, .
\end{equation}
Using the polarization relation for kinetic Alfv\'en waves (Equation~\eqref{eq:dn2_dB2_kin_alfven}), one finds

\begin{equation}\label{eq:Z2}
 Z^2_{d_r} = 
 \frac{\delta B^2 _{\perp {d_r}}}{{4\pi \, m_i \, n}} 
 =\beta_i \, (1+\beta_i) \, \frac{B^2}{4\pi \, m_i \, n} \, \left(\frac{d_r}{\rho_i}\right)^{\delta -1} \, 
 \frac{ \delta n^2 _{\rho_i}}{n^2}\,\,\, ,
\end{equation}
where we have used the $k_\perp^{-\delta}$ form of the spectrum in the kinetic range $k_\perp \geq d_r^{-1}$ (right panel of Figure~\ref{fig:obs_cartoon}) to scale the density fluctuation level $ \delta n^2 _{\rho_i}$ inferred from radio observations to its value at wavenumber $d_r^{-1}$. 

\begin{figure}
\centering
\includegraphics[width=0.8\textwidth]{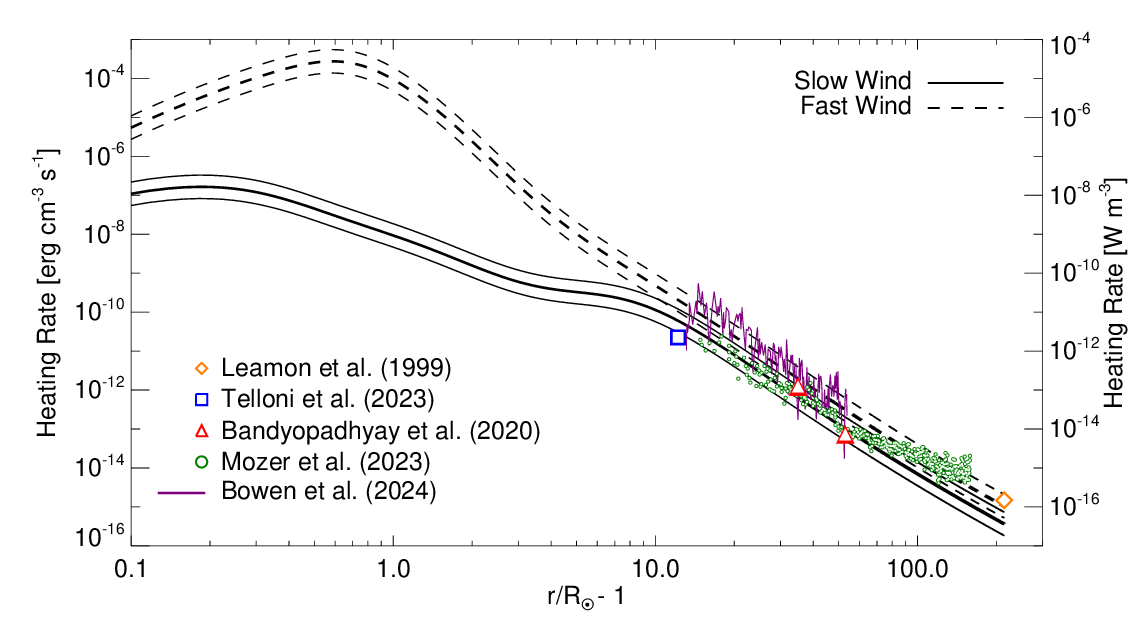}
\caption{\label{fig:Heating} Energy cascade rate (Equation~\eqref{eq:heating_rate}) using two models of the corona: an equatorial active region with slow solar wind (solid lines), and a coronal hole with fast solar wind (dashed lines). The bands represent a scaling factor ranging between $0.5$ and $2$. The overplotted data are from in-situ measurements by \cite{1999JGR...10422331L, 2020ApJS..246...48B, 2023ApJ...955L...4T, 2023AandA...673L...3M} and \cite{2024ApJ...972L...8B}. }
\end{figure}

Figure~\ref{fig:Heating} shows the radial profile of the heating rate using Equations~\eqref{eq:heating_rate} and~\eqref{eq:Z2}, where the profiles of density, magnetic field and temperature are given in Appendix~\ref{sec:plasma_params}, and the radio-inferred fractional density fluctuations $\delta n^2 _{\rho_i}/n^2$ at the ion gyroradius scale~$\rho_i$. The heating rate is lower in the corona for the slow solar wind (solid line) than for the fast solar wind (dashed line), with the largest differences occurring when $\beta_i \ll 1$, i.e., $d_r/\rho_i \simeq v_A/v_{Ti} \gg 1$. Equation~\eqref{eq:heating_rate} can be simplified to highlight the dependency on the plasma parameters:

\begin{equation}\label{eq:heating_rate2}
 \frac{Q (r)}{n \, k_B \, T_i} \propto \frac{\delta n_{\rho _i}^3}{n^3} \, \omega_{ci} \, \left(v_A/v_{Ti}\right)^{(\delta -5)/2} \propto \frac{\delta n_{\rho _i}^3}{n^3} B^{(\delta-3)/2} \, (n \, T_i)^{(5-\delta)/4} \,\,\, ,
\end{equation}
so that $Q (r) \propto (\delta n_{\rho _i}^3/n^3) \, B^{(\delta-3)/2} \, (n \, T_i)^{(9-\delta)/4}$, highlighting the strong dependency of the heating rate on the plasma density for a given level of density fluctuations. With $\delta \simeq 3$, the heating rate $Q \propto (nT_i)^{3/2}$ and is independent of $B$. For $\delta = 8/3$, $Q \propto B^{1/6} \, (nT_i)^{19/12} \propto B^3 \, \beta_i^{19/12}$. Thus, at a location with a given ion temperature $T_i$, higher values of both the density $n$ (or, equivalently, the ion plasma beta $\beta_i$) and the magnetic field $B$ lead to a higher heating rate. 

In the heliosphere, the cascade rate is estimated using in-situ measurements, which provides an opportunity to compare the remote and in-situ diagnostics over a wide range of distances measured at different inertial range scales. Overall, the cascade rate and the heating rate decrease with distance (Figure~\ref{fig:Heating}). For example, \citet{2020ApJS..246...48B} used data from the PSP FIELDS \citep{2016SSRv..204...49B} instrument, at heliocentric distances ranging from $\sim36$~R$_\odot$ to $\sim54$~R$_\odot$, to determine the energy transfer rate at a much larger scale around $500 \, d_r$, well within the inertial range. \citet{2023ApJ...955L...4T} estimated coronal heating rate in the slow solar wind by combining data from the Metis coronograph \citep{2020A&A...642A..10A} on the Solar Orbiter mission and PSP ion density and magnetic field measurements. \cite{2023AandA...673L...3M} used PSP measurements of proton distributions in the range from $\sim20$~R$_\odot$ to $\sim160$~R$_\odot$ to deduce a perpendicular proton heating rate (and also showed that parallel protons are neither heated nor cooled between $\sim20$~R$_\odot$ to $\sim70$~R$_\odot$, and ``most likely not heated or cooled beyond that distance''). \cite{2024ApJ...972L...8B} studied PSP FIELDS observations of the magnetic field fluctuations within an extended stream of fast solar wind ranging from $\sim15$~R$_\odot$ to $\sim55$~R$_\odot$. Figure~\ref{fig:Heating} summarizes the results of these cascade rate estimates superimposed on the KAW heating rate model of Equation~\eqref{eq:heating_rate}, deduced using the density fluctuation model from radio observations; a consistency is evident for the range of distances from the Sun where in situ measurements are available.

\section{Summary}\label{sec:summary}

The density fluctuations inferred from radio observations of the location, size, and decay times of type~III solar radio bursts are found to be consistent with 
that values expected from kinetic Alfv\'en waves or KAW structures over a broad range of distances. The magnetic fluctuations deduced using the radio measurements are similar to in-situ measured magnetic 
fluctuations at ion-scales in the inner heliosphere between $0.1-1$~au. 
Moreover, the density fluctuation amplitudes deduced using remote observations allow to deduce the radial variation of the kinetic Alfv\'en wave magnetic fluctuations at ion scales in the corona from $\sim 0.1$~R$_\odot$, where in-situ measurements are not available.

Using this result, we have estimated the turbulence energy cascade rate 
near ion scales (where the wave spectrum transitions from inertial to kinetic scales), and we find that the rate is very similar to the energy transfer 
rate obtained in the solar wind at larger inertial scales from 
in-situ measurements. 
The radio-inferred heating rate decreases with distance 
quantitatively similar to in-situ measurements 
reported in the literature. 
As the radio observations are available from close to the Sun from 
$\sim 0.1$~R$_\odot$ to 1~au, 
the energy cascade rate can be estimated at the distances
where in-situ measurements are absent. 
The heating rate is larger for fast solar wind parameters 
and weaker for the slow wind. 
The coronal hole (fast) solar wind heating reaches values
$10^{-5}$~Wm$^{-3}$ around $1R_\odot$ away from the Sun 
and could provide sufficient energy input to explain the observations.

In summary, as the density fluctuations are inferred over a wide range of distances 
from close to the Sun near $\sim 0.1$~R$_\odot$ to 1~au, 
the magnetic fluctuations and energy cascade rate can also be calculated over an unprecedented 
range of distances constraining coronal heating models. 
The equivalent energy flux supplied by the turbulent heating $\int _{r=1.1R_\odot}^{1 \, {\rm au}} Q(r) \, dr$ is $\sim$ is between $10^4$~erg~cm$^{-2}$~s$^{-1}$ ($10$~W~m$^{-2}$) and 
$10^7$~erg~cm$^{-2}$~s$^{-1}$ ($10^4$~W~m$^{-2}$) 
for the slow and fast solar wind, respectively.
The values are comparable \citep[e.g.,][]{1977ARA&A..15..363W,1986JGR....91.4111H,1988ApJ...325..442W} to the typical energy flux requirements to heat the corona. 

\begin{contribution}
This project was initiated by EPK and DLC. The development of the narrative was carried out principally by EPK and AGE, and the data and figures were produced primarily by DLC and AP. All authors contributed to the final version of the manuscript and agree with its content.
\end{contribution}

\begin{acknowledgments}
 
We thank the referee for some very insightful comments that have significantly improved the manuscript. EPK and DLC acknowledge financial support from the STFC/UKRI grant ST/Y001834/1. EPK is supported by the Leverhulme Trust (Research Fellowship RF-2025-357). AGE was supported by NASA's Heliophysics Supporting Research Program through award 80NSSC24K0244, by the NASA EPSCoR program through award number 80NSSC23M0074 to NASA Kentucky, and by the Kentucky Cabinet for Economic Development. AP acknowledges support from the MSMT grant
LUAUS25060. We also acknowledge support from the International Space Science Institute for the LOFAR \url{http://www.issibern.ch/teams/lofar/} team. This research has made use of NASA's Astrophysics Data System Bibliographic Services.

\end{acknowledgments}

\appendix

\section{Plasma density, magnetic field and temperature profiles}\label{sec:plasma_params}

Slow solar wind plasma parameters are taken to match PSP observations and provide typical parameters in the corona and inner heliosphere. The plasma temperature is given by

\begin{equation}\label{eq:T_r_slow}
 T(r)=\frac{3\times 10^6}{(r/R_\odot+1)^{0.7}} \;\;\;\text{[K]} \,\,\, , 
\end{equation}
where the $T(r) \sim r^{-0.7}$ dependence at large $r$ corresponds to the PSP observations. The plasma density is taken to be 

\begin{equation}\label{eq:n_r_slow}
 n(r)=\left[61.9\left(\frac{R_\odot}{r}\right)^{12.67}
 +16.4\left(\frac{R_\odot}{r}\right)^{5.14}\right] \times 10^7
 +10\left(\frac{215R_\odot}{r}\right)^{1.8}
 \;\;\;[\text{cm}^{-3}] \,\,\, ,  
\end{equation}
where the coronal part (the terms in the square brackets) corresponds to the model of \citet{1996ApJ...458..817G} and the remaining term is a fit to the PSP measurements in the heliosphere. Note that this density profile is similar to the one deduced by \citet{2025ApJ...979L..10K}. 

The magnetic field strength is given by

\begin{equation}\label{eq:B_r_slow}   
 B(r)=0.5 \, \left(\frac{r}{R_\odot}-1\right)^{-1.5} \, \left(\frac{r}{10 \, R_\odot}+1\right)^{-2} \, 
 +3\times10^{-5}\left(\frac{215 \, R_\odot}{r}\right)^{1.8} \text{[G]} \,\,\, ,
\end{equation}
where the first first term is given by \cite{1978SoPh...57..279D} for near-solar distances $r < 10$~R$_\odot$ and the second term is the interplanetary magnetic field consistent with the measurements by PSP (Figure~\ref{fig:psp_plasma}).

For the fast solar wind we adopt coronal hole parameters with a lower coronal 
temperature around $1.5$~MK:

\begin{equation}\label{eq:T_r_fast}
 T(r)=\frac{3\times 10^6}{(r/R_\odot+1)^{0.7}} \, 
 \frac{r/R_{\odot}}{(r/R_\odot+1)} \;\;\;\text{[K]} \,\,\, ,
\end{equation}
a smaller number density

\begin{equation}\label{eq:n_r_fast}
 n(r)= \left[ 1737 \left( \frac{R_\odot}{r} \right)^{14}
 +20 \left( \frac{R_\odot}{r} \right)^{4.1}+2\left( \frac{R_\odot}{r} \right)^2\right] \times 10^5
 \;\;\;[\text{cm}^{-3}]
\end{equation}
\citep{1999JGR...104.9801G}, and a magnetic field 

\begin{equation}\label{eq:B_r_fast}   
 B(r) = 12 \, \left( \frac{R_\odot}{r} \right)^{3.5} + 1.5 \, \left( \frac{R_\odot}{r} \right)^{2} \, \text{ [G]}
\end{equation}
\citep[Equation~(7) in][]{1999JGR...10424781H}, which gives a magnetic field $\simeq$$13$~Gauss in the low corona.

\bibliography{refs}{}
\bibliographystyle{aasjournalv7}

\end{document}